\documentclass[useAMS,usenatbib]{mn2e}
\usepackage{graphicx,epsfig,amssymb,amsmath,layout,verbatim,rotating,calc,mathrsfs,natbib}
\usepackage{graphics}
\usepackage{rotate}
\usepackage{txfonts}

\newcommand{\Hi}{{\rm H{\small I }}}

\title[The NGC~1300 potential]{NGC~1300 Dynamics:\\ I. The gravitational
  potential as a tool for detailed stellar dynamics}
\author[C.~Kalapotharakos et al.]
{C.~Kalapotharakos,$^{1}$\thanks{ckalapot@phys.uoa.gr
  (CK); patsis@academyofathens.gr (PAP); pgrosbol@eso.org (PG)}
  P.A.~Patsis,$^{1,2,3}$ and P.~Grosb{\o}l$^{3}$\footnotemark[1]\thanks{Based
    on observations collected at the European Southern
Observatory, Chile: program: ESO 69.A-0021.}\\
$^1$Research Center for Astronomy, Academy of Athens, Soranou Efessiou
    4, GR-115 27, Athens, Greece\\
$^2$ Observatoire Astronomique de Strasbourg, 11 rue de l'Universit\'{e},
    67000 Strasbourg, France\\
$^3$ European Southern Observatory, Karl-Schwarzschild-Str. 2, 85748 Garching,
    Germany }

\date{Accepted ..........Received .............;in original form ..........}

\begin{document}

\maketitle

\label{firstpage}

\begin{abstract}
In a series of papers we study the stellar dynamics of the grand
design barred-spiral galaxy NGC~1300. In the first paper of this
series we estimate the gravitational potential and we give it in a
form suitable to be used in dynamical studies. The estimation is
done directly from near-infrared observations. Since the 3D
distribution of the luminous matter is unknown, we construct three
different general models for the potential corresponding to three
different assumptions for the geometry of the system, representing
limiting cases. A pure 2D disc, a cylindrical geometry (thick
disc) and a third case, where a spherical geometry is assumed to
apply for the major part of the bar. For the potential of the disc
component on the galactic plane a Fourier decomposition method is
used, that allows us to express it as a sum of trigonometric
terms. Both even and odd components are considered, so that the
estimated potential accounts also for the observed asymmetries in
the morphology. For the amplitudes of the trigonometric terms a
smoothed cubic interpolation scheme is used. The total potential
in each model may include two additional terms (Plummer spheres)
representing a central mass concentration and a dark halo
component, respectively. In all examined models, the relative
force perturbation points to a strongly nonlinear gravitational
field, which ranges from 0.45 to 0.8 of the axisymmetric
background with the pure 2D being the most nonlinear one. The
force perturbation in each model is found being robust to small
changes of the required parameter values. We present the
topological distributions of the stable and unstable Lagrangian
points as a function of the pattern speed $(\Omega_p)$. The
topological distribution found deviates in several cases from the
classical paradigm with two stable Lagrangian points at the sides
of the bar and two unstable ones close to the ends of the bar. In
all three models there is a range of $\Omega_p$ values, where we
find multiple stationary points whose stability affects the
overall dynamics of the system.
\end{abstract}



\begin{figure*}
\begin{center}
\includegraphics[width=\textwidth]{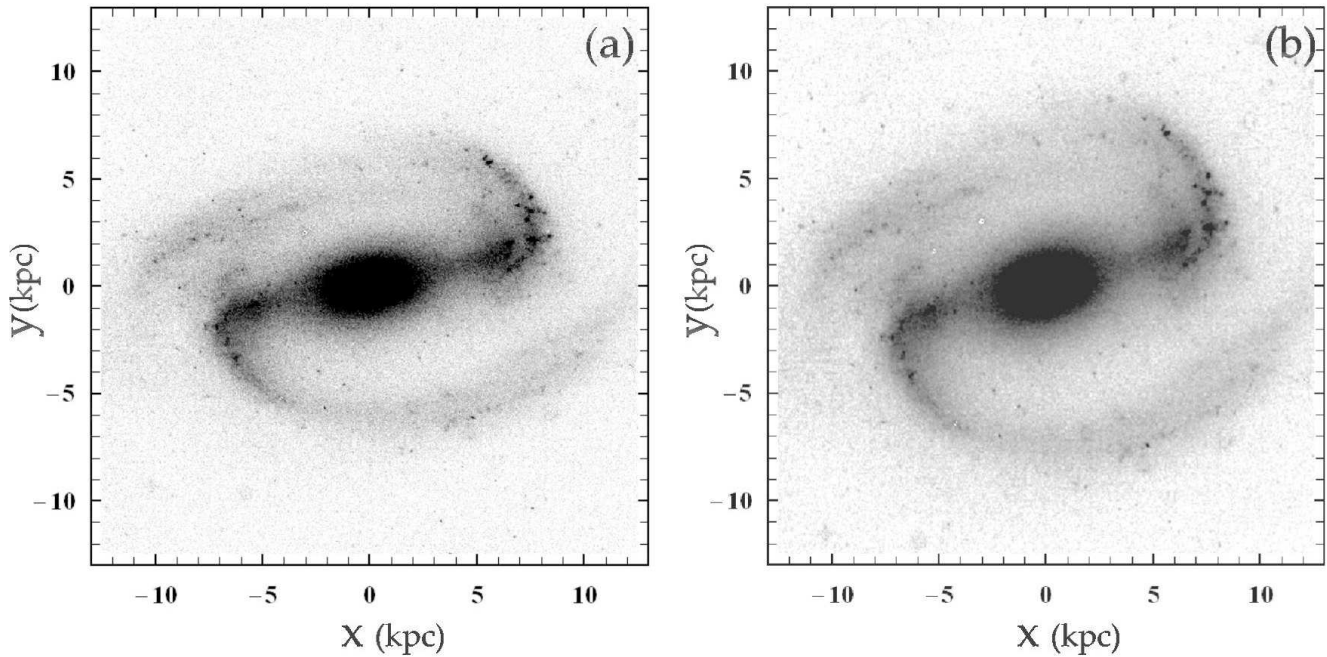}
\end{center}
\caption{(a) The observed K band image of NGC~1300.
 (b) The deprojected K band image of NGC~1300 using (PA,
IA)=$(87^{\circ}, 35^{\circ})$.  The galaxy was observed with SOFI
at the 3.5~m NTT telescope at ESO La Silla. The units on the axes
are in kpc, assuming a distance to the galaxy $D=19.6$Mpc. The
images are in linear intensity scale.} \label{fig01}
\end{figure*}

\section{Introduction}
Grand design galaxies are attractive objects for dynamical
studies, because it is believed that behind their coherent global
features are hidden the basic dynamical mechanisms that shape the
morphology of galactic discs. Then, deviations from the grand
design could be considered to reflect additional complications of
this basic dynamical behavior. The result would be more
complicated morphologies, like those encountered in multi-armed or
flocculent spiral galaxies. The motivation of this study is to
investigate the dynamics of NGC~1300, which is a well known
barred-spiral grand design galaxy with a prominent bar. Its
archetypical morphology promises a deeper understanding of the
dynamics of a class of similar objects.

Barred-spiral systems offer many examples of grand design
morphology, being characterized by a set of two spiral arms starting
close to the ends of the bar. In some cases the spirals seem to
emerge out of the bar as its continuation, while in others the
beginning of the spirals is clearly displaced from the bar's ends
\citep[e.g. compare the spirals of NGC~4535 and NGC~4548 in figure 5
in][]{gro2008}. Nevertheless the morphology of the arms in many of
these barred-spiral systems is not as smooth as that of the arms of
normal spiral galaxies \citep{gropat1998}, especially in
near-infrared wavelengths. This can be characteristically seen by
overplotting the $m=2$ component of the surface brightness on images
of galaxies, as do e.g. \citet{sea2005}. One can observe that
their arms are in general asymmetric and have gaps. It is also clear
that in many cases the surface brightness of the arms is larger
close to the ends of the bar than at azimuths away from them.
Characteristic cases are NGC~4314 and NGC~4665 \citep{gad2008}. In
the case of NGC~4314, in near-infrared wavelengths, the arms hardly
complete azimuthally a $\pi/2$ angle \citep{quietal1994}.

In the last years the dynamics of barred-spiral systems have been
revisited by several research groups. The reason is the
investigation of the possibility that the stars on the spirals are
in chaotic motion \citep[see also articles by the same authors in][and
  references
therein]{cia2008}. The motivation for this series of papers is
exactly to test this hypothesis in one of the well known and well
studied galaxies of this type, NGC~1300. The first step is to
estimate its potential directly from observations
\citep{devacetal1991}. Since it is a well studied object, one can
find observations from X-rays to radio wavelengths in order to
understand the detailed morphology \citep[e.g.][]{fgt88, e89}. The
composite color image STScI-PRC2005-01
(http://heritage.stsci.edu/2005/01/supplemental.html), highlights
the characteristic asymmetry between the two sides of the galaxy.

Images in optical wavelengths, as well as B$-$I color maps that
cover an area of radius $\approx6\arcmin$ from the centre of the
galaxy, reveal faint extensions of the main bisymmetric spiral
structure \citep[e.g.][]{eech96}. These extensions, more clearly
observed in bluer bands, consist of
Population I objects.

In Fig.~\ref{fig01}a we present the K band image of the galaxy, that
we used in the present study. Images at near-infrared wavelengths
depict the morphology of the old disc stellar population and are
appropriate for dynamical studies (see Sect.~\ref{obs} below). The
axes labels are in kpc, assuming a distance to the galaxy
$D=19.6$~Mpc. At this distance 1\arcsec $\approx0.095$kpc.

Our ultimate goal is to present dynamical mechanisms for the
stellar component that lead to the development of the particular
morphological features of the galaxy NGC~1300. The estimation of a
suitable potential is the basis for all dynamical studies and this
is done in the present paper, which is structured as follows: In
Section 2, we present the observations we performed in order to
obtain a reliable estimation of the potential of the galaxy. In
Section 3 we describe the image processing techniques that allowed
us to create images that will be used to compare our models with.
The estimation of the potential is given in Section 4. The forcing
in the models is presented in Section 5, while the properties of
the corresponding effective potentials in Section 6. Finally we
discuss our conclusions in Section 7.

\section{Observations}\label{obs}
The distribution of luminous matter in a galaxy is indicated by its
surface brightness map. The translation from luminosity to mass,
through a mass-to-light ratio $(M/L)$, may vary significant
depending on both the spectral band used and the underlying stellar
population. The near-infrared K band at 2.2~$\mu$ is close to the
emission peak of the old stellar disc and bulge populations which
represent the major visual mass constituency of a spiral galaxy
\citep{rixrie1993}. Although very young stars and red super giants
also contribute to the K band flux, it remains a good indicator of
the mass distribution of luminous matter outside star forming
regions.

The K-map of NGC~1300 was observed on 2002-09-01 with SOFI at the
3.5~m NTT telescope, ESO La Silla. The total exposure time on
target was 10~min in the K$_\mathrm{s}$ filter, at 2.162$\mu$,
with equal time for sky fields. A jitter pattern with $10\arcsec$
shifts was used to allow removal of bad pixels through stacking,
while sky frames were interleaved with offsets around $10\arcmin$
from the galaxy. The SOFI instrument had a Rowckwell Hg:Cd:Te
1024$\times$1024 detector with $0.29\arcsec$ pixels on the sky.
The field covered the main spiral structure of NGC~1300 but did
not provide an accurate estimate of the sky background level. The
final K band frame had a seeing of 0.8\arcsec and reached a
surface brightness of \mbox{K = 20.8 mag \text{arcsec}$^{-2}$} at
a signal-to-noise (S/N) level of 3. Thus, the quality of our
dedicated SOFI/NTT observations can be considered as an additional
advantage for potential calculations, besides the known advantages
of the K-band for studies related with the mass distribution in
galaxies. For comparison, the 2MASS survey offers a seeing of
2-3\arcsec and the corresponding images reach
20~mag~\text{arcsec}$^{-2}$ at a S/N  level of 1$\sigma$ (see
pages under http://www.ipac.caltech.edu/2mass/). The reduction of
the SOFI data followed the procedure described by \citet{gpp2004}.


The sky projection parameters were adopted from \citet{linetal1997}
who estimated the Position (PA) and Inclination Angle (IA) by
fitting tilted ring models to their \Hi velocity data. The
parameters values they found were (PA, IA) = $(87^\circ, 35^\circ)$.
These values are different from those one finds by fitting an
exponential disc to the regions outside the main bar in the K band
image, as in \citet{gpp2004}, which are (PA, IA)=(106.6$^\circ,
42.2^\circ$). This values are close to those found by \citet{ls02}.
However, since for barred spirals like NGC 1300, any determination
of the sky projection based on surface photometry may be strongly
biased due to the strong $m=2$ mode and the open spiral structure,
we preferred to use for our study the values proposed by
\citet{linetal1997}. A disadvantage of the use of the near-infrared
image is that one can hardly detect the axisymmetric disc in the
outer regions. We have to note though, that for the response models
and the orbital analysis, we investigated also the effect of the
deprojection parameters in the potential, and the results are
presented in the forthcoming papers of this series.

The deprojected with parameters (PA, IA) = $(87^\circ, 35^\circ)$
image of NGC~1300, is shown in Fig.~\ref{fig01}b and depicts the
morphology we will try to model. The size of this image is
531$\times$531 since it has been re-binned to 0.5\arcsec pixel size.
The standard background subtraction for the K-band frame could be
slightly in error since it was based on separate sky exposures and
not on the frame itself which was almost fully occupied by the
galaxy.

\begin{figure*}
\begin{center}
\includegraphics[width=15cm]{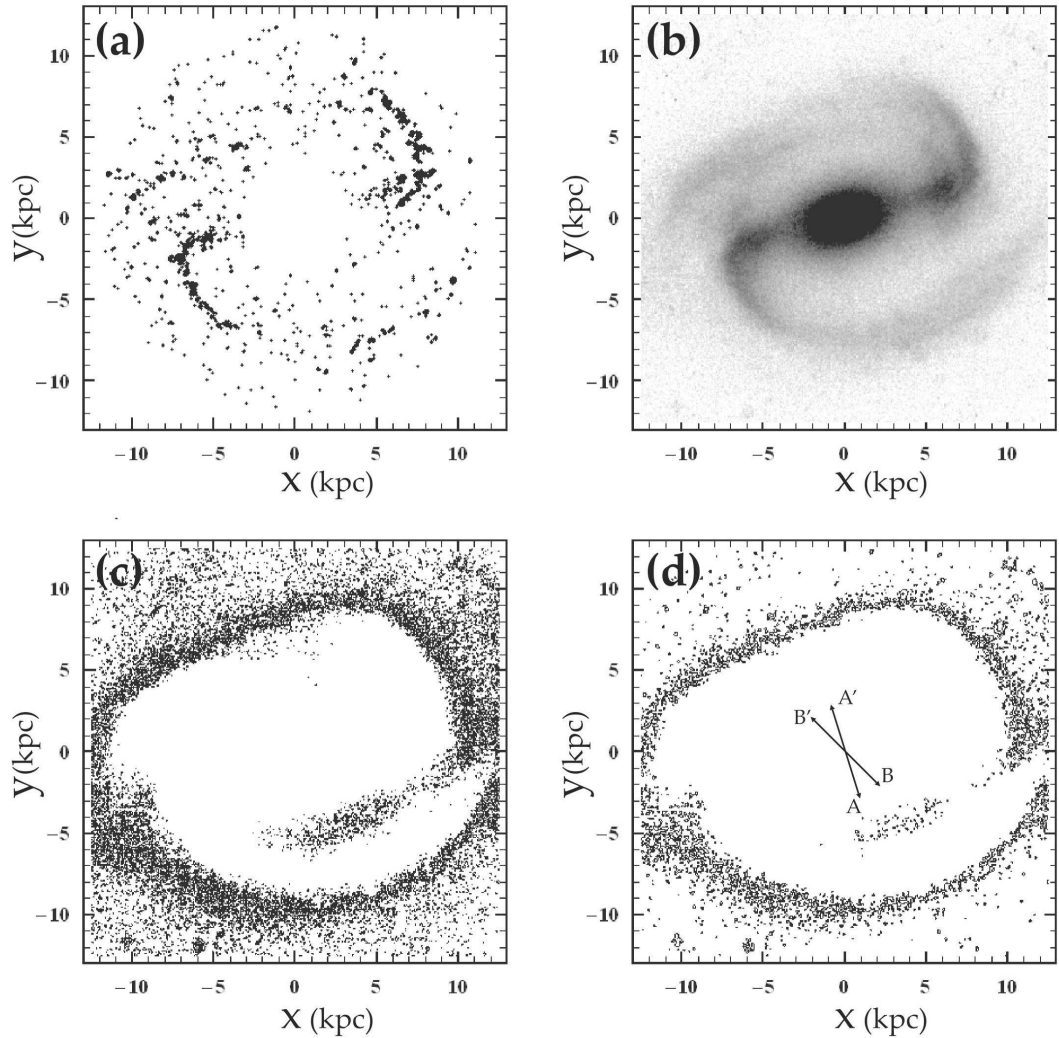}
\end{center}
\caption{(a) The bright spots identified as young clusters.
These regions correspond to extreme high valued pixels. (b)
The K band image of NGC~1300 after reducing the intensities of the
extreme high valued pixels, This image is smoother than the one in
Fig.~\ref{fig01}b and lacks of bright areas corresponding to young
clusters. (c) The contour line for 20.8mag~arcsec$^{-2}$
for the image panel (b). It is obvious that the low brightness data
are noisy. (d) The same contour line, as in (c), after the
application of a low-pass smoothing filter to low brightness pixels
$(\gtrsim 19\text{mag~arcsec}^{-2})$. The arrows A, A', B, B' denote
the directions along which we plot the surface brightness in
Fig.~\ref{fig04}.} \label{fig03}
\end{figure*}

\section{Data reduction}
We adopted as distance of the galaxy the value $D=19.6~\text{Mpc}$,
according to the cosmology-corrected luminosity distance given in
the NASA/IPAC Extragalactic Database (NED). This value may differ
from the distance used in other papers on the dynamics of NGC~1300.
In any case, the exact distance of a galaxy does not influence
essentially the dynamical phenomena that shape its morphology
\citep{pcg91}.

Before the K-band frame can be used to bootstrap the estimate of
the potential, two possible corrections should be considered
namely the population effect on the $(M/L)$ ratio and the exact
background level.  Although the major fraction of flux in the
K-band originates from the old stellar, disc population
\citep{rixrie1993}, young stars in the arm regions will bias the
effective $M/L$. The young stellar population may be present both
as a diffuse, unresolved, component and as resolved young stellar
complexes seen in the arms as bright knots \citep{phg01,gd08}. In
order to reduce the $M/L$ bias from young stars, we removed such
bright knots by a digital filter. A correction that has to be done
concerns  the brightness of the outer parts of the galactic disc.
This area is biased towards higher values due to the presence of
young stellar clusters and stellar complexes, still present even
in K band images \citep{phg01, gd08}. For this purpose, in
Fig.~\ref{fig01}b we firstly removed the bright spots in the
region with a radius $R>6.5\text{kpc}$. Bright spots, assumed to
be young clusters, do not cover areas larger than $10\times 10$
pixels. We substituted the intensities of all extreme high (but
also a few low) valued pixels with the median value of the
corresponding $961$  $(31\times 31)$ pixels area centred at the
very pixel under consideration.
In Fig.~\ref{fig03}a  only the
 identified as high valued pixels are shown, while in
Fig.~\ref{fig03}b we see the rectified image of the galaxy after the
removal of the bright spots.
After these reductions the galaxy image seems quite smooth in the
area of high valued pixels (Fig.~\ref{fig03}b) but still presents
significant noise in the area of low valued pixels. In Fig.~\ref{fig03}c we see
how noisy is the contour corresponding to the brightness value
$20.8\text{mag~arcsec}^{-2}$. In order to smooth out the image, we
apply a low-pass smoothing filter which affects only the low
valued pixels ($\gtrsim 19\text{mag~arcsec}^{-2}$) getting a less
noised image (Fig.~\ref{fig03}d).

\begin{figure*}
\begin{center}
\includegraphics[width=15cm]{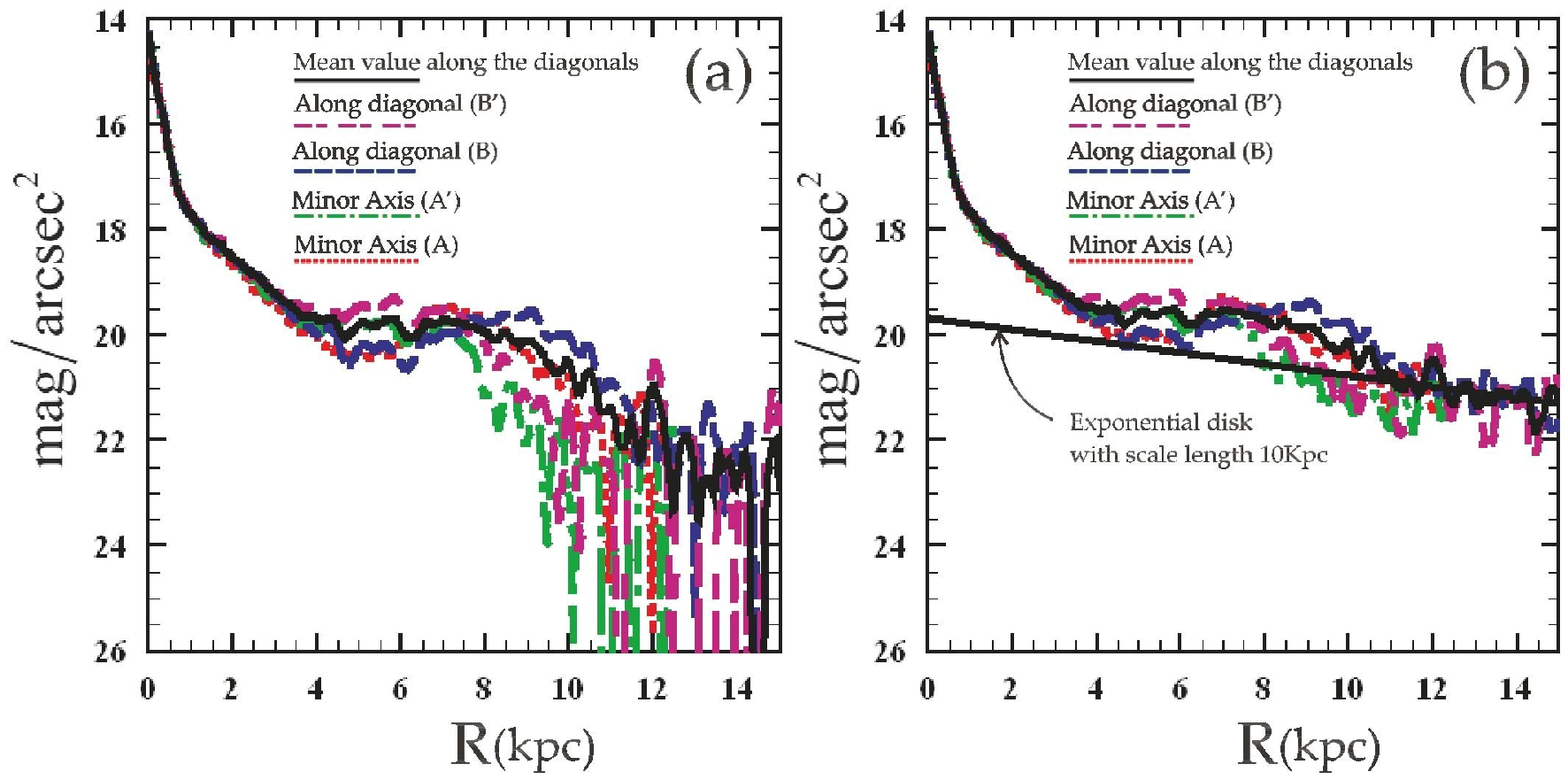}
\end{center}
\caption{(a) The surface brightness along the directions
indicated in Fig.~\ref{fig03}d. We observe that no exponential law
can be identified in the outer parts, which means that the
considered zero level is not correct. (b) The surface
brightness along the same directions after the minimum correction of
the zero level so that an exponential decrement is obtained in the
outer parts. The scale length of the resulting exponential law is
10~kpc.} \label{fig04}
\end{figure*}

In Fig.~\ref{fig03}d we indicate four radial directions A,
A$^\prime$, B and B$^\prime$, centred at the axes origin, along
which we plot the surface brightness $\Sigma(R)$ after applying the
smoothing filter. We have chosen the directions along the minor axis
(A, A$^\prime$) which reveal better the exponential disc. The
directions (B, B$^\prime$) are along the diagonal of the image
nearest to the minor axis and provide data for longer distance. The
surface brightness along these directions is given in
Fig.~\ref{fig04}a. It is obvious that no exponential profile is
observed at the outer parts of the disc (beyond the region with the
bar and spiral structures) as expected. This implies that the
adopted zero value level for the background is not correct.
Therefore, we decrease the zero value level by the smallest quantity
so that the surface brightness profiles become exponential, i.e.
\begin{equation}\label{sigexp}
    \Sigma(R)\propto e^{-R/h_R},
\end{equation}
as shown in Fig.~\ref{fig04}b (actually we increase the brightness
values of all pixels by the same quantity). We consider the region
with the spiral arms part of the exponential disc. We note that
the applied correction is of the order of the background
uncertainty. The corresponding scale length we find in that case
is $h_R\simeq 10~\text{kpc}$. By increasing the brightness by a
smaller value than the one we finally applied, we do not get a
well defined exponential profile. The value of 10~kpc we find is
in agreement with the one given by \citet{pea2001} corresponding
to observations of NGC~1300 in the I band. We note that
\citet{ls02} and \citet{lsbv04} find values close to 7~kpc in
their 2MASS and H-band images respectively. The effect of such
differences on the nonlinearity of models for the potential is
investigated below (Sect.~5).

\begin{table*}
\caption{All the parameter values used in \eqref{pottot},
\eqref{potcmcdh} and \eqref{potmhp} for each model (see text). These
values correspond to the rotation curves of the models in Figs.~\ref{rotcur_2D}, \ref{rotcur_3Dbulgedisk} and \ref{rotcur_3Ddisk}.
The last two columns read the total mass inside the (spherical) radius $r=15$kpc
for the luminous (LC) and dark (DC) component respectively.}
\centering \label{tabt000}
\begin{tabular}{@{}lcccccccc@{}}\hline 
  models & $M_{CMC} (M_{\odot})$ & $a_{CMC}$ (kpc) & $M_{DH} (M_{\odot})$ & $a_{DH}$ (kpc) & $j_0 (M_{\odot}\text{kpc}^{-3})$ & $a_{MHP}$ (kpc) & LC $(M_{\odot})$ & DC $(M_{\odot})$ \\ \hline
  Model A & $7.0\times 10^8$ & 0.1 & $8.0\times 10^{11}$ & 30 & - & - & $5.0\times 10^{10}$ & $7.6\times 10^{10}$ \\ 
  Model B & $2.3\times 10^9$ & 0.1 & $1.6\times 10^{12}$ & 40 & - & - & $6.5\times 10^{10}$ & $7.5\times 10^{10}$  \\ 
  Model C & $9.4\times 10^8$ & 0.07 & $1.6\times 10^{12}$ & 40 & $6.12\times 10^{10}$ & 0.19 & $7.1\times 10^{10}$ & $7.5\times 10^{10}$  \\ \hline
\end{tabular}
\end{table*}

\section{The estimation of the potential}
After all the above amendments we result in an estimation for the
surface density distribution of the \textit{luminous} component in
the reduced image (Fig.~\ref{fig03}b), under the assumption of a
constant mass-to-light ratio. In our approach we reconstruct the
galactic potential assuming the maximum contribution of the
luminous component to the observed rotation curve as given in
\citet{linetal1997}. For the cases that parts of the rotation
curve will not be able to be entirely assigned to the luminous
component we switch on two additional components in our modelling.
They represent a dark halo and/or a central mass concentration.
Since our goal is the study of the dynamics of the stars, treated
as test particles moving in the estimated gravitational field, the
total potential must be compatible with the rotation curve of the
galaxy, in every model we study.

In order to proceed with the potential estimation of the luminous
component we need to do an assumption about the distribution of
the matter in the third dimension. Below, we study three different
cases. The basic difference among these three general models is
their \textit{geometry}. We want to compare the marginal thin
geometry of the 2D case (model A) with a thick disc case (model B)
in order to see if the comparison between galaxy and models
improves by varying our parameters towards the one or the other
direction in the parameter space. Model C changes essentially the
geometry of the bar from cylindrical to a combination of spherical
with cylindrical and should be again considered as a limiting
case. More specifically we consider:

\begin{description}
    \item[\textbf{Model A.}] The `2D' or `degenerate' case in which all
matter is considered lying on the ($x,y$) galactic plane.
    \item[\textbf{Model B.}] The `disc' case in which we assume a cylindrical
      geometry and that all matter is
distributed in a 3D disc with scale height $h_z$. Then the
vertical dependence of the density $\rho$ reads
\citep[][]{kruit1988}
\begin{equation}\label{rhoz}
    \rho\propto\rho_z(z)=\dfrac{1}{2h_z}\text{sech}^{2/n}\left(\dfrac{n z}{2
      h_z}\right),
\end{equation}
where $h_z$ is the vertical scale height and $n$ is an index (see
Section 4.2, for its meaning).
    \item[\textbf{Model C.}] The `spheroidal' case in which we
      consider that the total observed brightness is the combination
      of two major components: i) a spheroidal component that accounts for the
      major part of the bar and ii) a 3D exponential disc with scale height
      $h_z$ as in model B. The spheroidal component includes the
      bulge \citep[$r<1.5$kpc,][]{pea2001,ls02,lsbv04} and the
      axisymmetric contribution of the central part of the bar.
\end{description}

In Cartesian coordinates for a given spatial density $\rho(x,y,z)$
the gravitational potential $\Phi(x,y,z)$ is
\begin{equation}\label{gravpot}
    \Phi(x,y,z)=-G\int
    \frac{\rho(x',y',z')dx'dy'dz'}{\sqrt{(x-x')^2+(y-y')^2+(z-z')^2}},
\end{equation}
where $G$ is the gravitational constant. This means that
$\Phi(x,y,z)$ is the convolution of the functions $\rho(x,y,z)$
and $g(x,y,z)=\dfrac{1}{r}=\dfrac{1}{\sqrt{x^2+y^2+z^2}}$. Thus
the potential is written as $\displaystyle{\Phi=\rho\otimes g}$,
where $\otimes$ denotes the convolution operator.

\subsection{Model A}
In model A, where all mass is assumed on the $(x,y)$ plane, the
density can be written as
$\displaystyle{\rho(x,y,z)=\Sigma(x,y)\delta(z)}$, where $\Sigma$
is the surface density and $\delta$ is the delta function.
Substituting the above density expression into \eqref{gravpot} and
integrating it over $z$ we get
\begin{equation}\label{potA}
     \Phi(x,y,0)=-G\int
    \Sigma(x',y') g(x-x',y-y',0) dx'dy'
\end{equation}
which is the convolution of the functions $\Sigma(x,y)$ and
$g(x,y,0)$. Knowing the surface density
$\Sigma(x_i,y_j)$ at each grid point $(i,j)$ we can calculate
the potential at the same grid points by using the convolution
theorem
\begin{equation}\label{potconvth}
    \Phi=\mathcal{F}^{-1}\left[\mathcal{F}(\Sigma)\;\mathcal{F}(g)\right],
\end{equation}
where $\mathcal{F}$ and $\mathcal{F}^{-1}$, denote the Fourier
transform and inverse Fourier transform respectively. We use an
extended square area 16 times larger than the original area of the
image of the galaxy in order to avoid the contribution from
periodic reflections of the signal from the galaxy's pixels. Such
an extended area will allow us to consider, in dynamical studies,
stars that travel to large distances away from the centre of the
galaxy. The surface density in the extended area is considered to
be zero. We use the 2D fast Fourier transform (FFT) technique for
the corresponding computations. Note, that in this
calculation the adopted $g(0,0,0)$ value is equal to $(\text{bin
size})^{-1}\simeq 21~\text{kpc}^{-1}$ since the softening length is of the
order of the bin size. A detailed description of the whole
procedure can be found in \citet{hohhoc1969}.

Given the potential values at the grid points one can calculate
the potential everywhere in between these points using a 2D
interpolation scheme. However, such a scheme, besides its
complexity, provides the potential taking into account non-smooth,
local density disturbances. For this reason we proceed as in
\citet{quietal1994} and we decompose the potential into its
azimuthal Fourier components at every radius $R$. Under this
consideration the potential can be written in polar coordinates
$(R,\theta)$ as
\begin{equation}\label{potrthfour}
    \Phi(R,\theta)=\Phi_0(R)+\sum_{k=0}^{k_{max}}\left[\Phi_{kc}(R)\cos(k\theta)+\Phi_{ks}(R)\sin(k\theta)\right].
\end{equation}
For $k_{max}\rightarrow \infty$ the form \eqref{potrthfour}
returns exactly the same values with those we have from the 2D
interpolation scheme mentioned above. We cut off the high
harmonics and consider all terms (even and odd) up to $k_{max}=6$.
Like this we are able to reproduce all the important morphological
features of the galaxy (its asymmetries included).
\citet{quietal1994} calculated the amplitudes $\Phi_{kc}(R)$,
$\Phi_{ks}(R)$ for many rings $R_i$ and fitted them by 8th order
polynomials. Despite the advantages of simplicity and analyticity
of the expressions for each function that the use of polynomials
secures, it is not capable of following the actual variations
well. For this reason in the present study we apply the
interpolation scheme with cubic splines to the
$\left(R_i,\Phi_{kc}(R_i)\right)$,
$\left(R_i,\Phi_{ks}(R_i)\right)$ data sets. This scheme
guarantees the continuity of the potential, of the forces and of
derivatives of the forces, which are used in the variational
equations providing indicators that will measure the chaoticity of
the orbits (e.g. Lyapunov exponents). This is particular important
for stellar dynamical investigations, which is our motivation for
the estimation of the NGC 1300 potential. Note that we set
$\Phi_{kc}(0)=\Phi_{ks}(0)=0$ in order to secure the continuity
required at $R\rightarrow 0$. Table 2 in
Appendix\footnote{Appendix is only available in electronic form}
presents all the amplitude values, at 101 successive radii $R_i$,
used for the derivation of the interpolating splines.

\begin{figure}
\begin{center}
\includegraphics[width=8.4cm]{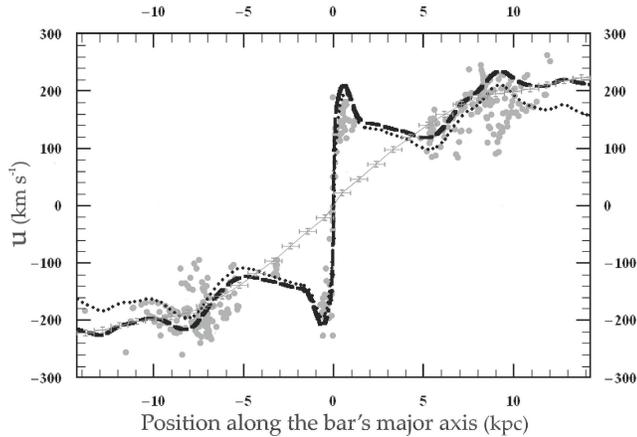}
\end{center}
\caption{The rotation curve of model A together with observational
data by \citet{linetal1997}. Gray dots correspond to optical
measurements from various slits aligned close to the major axis of
the galaxy. The gray curve with the error bars correspond the \Hi
rotation curve given in \citet{linetal1997}. The dotted line
represents the rotation curve of the luminous component only along
the major axis of the bar in model A. The dashed curve is obtained
by considering also an additional central mass (CMC) and a dark
halo (DH) component. The importance of these terms are determined
by the parameters given in Table 1.} \label{rotcur_2D}
\end{figure}

Figure \ref{rotcur_2D} shows the \Hi rotation curve (gray solid
line with error bars together with optical velocity measurements
(gray dots) from various slits aligned close to the major axis of
the galaxy, as taken from \citet{linetal1997}. In this figure we have also
plotted a rotation curve along the major axis of the bar, corresponding to the
potential
\eqref{potrthfour} of the luminous matter only (dotted line). In this case it
has
been considered being $5.0\times 10^{10} M_{\odot}$
(see Table \ref{tabt000}).
This curve fails reproducing the flat part of the rotation curve
at large distances. This implies that some additional mass
component exists besides the luminous component. The most
reasonable candidate to be invoked is a dark halo (DH) term. As
already mentioned our general model foresees the switching on of
two additional components for a central mass concentration (CMC)
and a DH. We model both of them with a Plummer sphere since it is
simple, effective and widely used in the literature. The
importance of this terms are controlled by the parameters of the
Plummer spheres. Thus the total potential $\Phi_T(R,\theta)$ can
then be written
\begin{equation}\label{pottot}
    \Phi_T(R,\theta)=\Phi_{LM}(R,\theta)+\Phi_{CMC}(R)+\Phi_{DH}(R)
\end{equation}
where $\Phi_{LM}$ is the luminous matter component given by
\eqref{potrthfour} and
\begin{equation}
\label{potcmcdh}
\Phi_{\begin{subarray}{l}
CMC\\ DH \end{subarray}}(R)=-\frac{G M_{\begin{subarray}{l} CMC\\
DH
\end{subarray}}}{\sqrt{R^2+a^2_{\begin{subarray}{l} CMC\\ DH
\end{subarray}}}},
\end{equation}
is the potential for the Plummer spheres, with $a$ being a constant parameter.
The adopted parameters are given in Table \ref{tabt000} (Model A).
In Fig.~5 we see that the DH term is quite important and helps the
rotation curve of the model reaching the level of the \Hi curve of
\citet{linetal1997}. On the other hand the CMC term is less
important increasing the central peak by $\approx12$km/sec which
eventually reaches the value $\approx208$km/sec. Note that the
central peak value is not well defined by the observational data
and therefore the adopted parameter values for the CMC term
provide only a reasonable area in the parameter space.

\subsection{Model B}

In case B we assume that the density can be written as
\begin{equation}\label{denscaseb}
    \rho(x,y,z)=\Sigma(x,y)\rho_z(z).
\end{equation}
\citet{kruit1988} proposed a family of models regarding the
vertical dependence of the density which reads
\begin{equation}\label{kruitmodel}
    \rho_z(z)=A\text{sech}^{2/n}\left(\frac{nz}{2h_z}\right)~~~\text{with}~~~
    n\in[1,\infty),
\end{equation}
where the normalization constant $A$ is given as
\begin{equation}\label{normconst}
    A=\frac{\Gamma\left(\frac{1}{2}+\frac{1}{n}\right)}{2h_z\sqrt{\pi}\;\Gamma\left(1+\frac{1}{n}\right)},
\end{equation}
so that 
$\int_{-\infty}^{\infty}\rho_z(z)dz=1$. $\Gamma$ is the Gamma
function defined by
\begin{equation}
    \label{gamma}
    \Gamma(w)=\int_{0}^{\infty}t^{w-1}e^{-t}dt.
\end{equation}

Varying  $n$ from $n=1 \text{ to } n\rightarrow\infty$ we get
$\rho_z(z)$ varying from
\[\rho_z(z)\propto \text{sech}^2\left(\frac{z}{2h_z}\right)\]
(isothermal disc) to \[\rho_z(z)\propto
\exp\left(-\frac{|z|}{h_z}\right)\] (exponential disc).
\citet{degrijs1998} and \citet{kreetal2002} provide estimations
for the mean values of the parameter $n$ and the ratio $h_R/h_z$
in samples of galaxies, which are $n\approx 3.7\pm 0.6$ and
$h_R/h_z\approx 7.3\pm 2.2$, respectively \citep[for a review
see][]{kruit2002}. These studies have been based on observations
in I and K bands.

\begin{figure}
\begin{center}
\includegraphics[width=8.4cm]{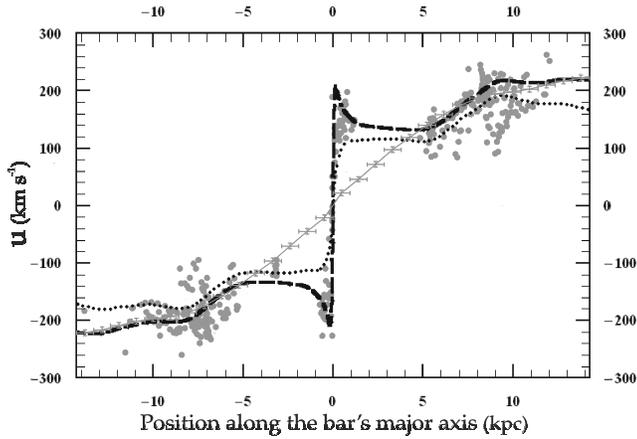}
\end{center}
\caption{The rotation curve along the major axis of the bar for
model B. The drawn curves are as in Fig.~\ref{rotcur_2D}. We
observe that the luminous component alone (dotted curve) reaches
lower velocity values at the central and outer parts of the galaxy
with respect to the velocity measurements. The added CMC component
is $\approx 3$ times more massive than in case A. The rotation
curve corresponding to the total potential \eqref{pottot} (dashed
curve) with the parameter values shown in Table \ref{tabt000}
levels off at about the same $u_{max}$ as the \Hi rotation curve.
Note that the central peak is sharper in this case.}
\label{rotcur_3Ddisk}
\end{figure}

We adopted the values $n=4,\;h_R/h_z=7$, which give $h_z\approx
1.43~\text{kpc}$ and following \citet{quietal1994}, we substitute
the density expression \eqref{denscaseb} into \eqref{gravpot} and
by integrating over $z$ we obtain
\begin{equation}\label{potB}
    \Phi(x,y,0)=-G\int
    \Sigma(x',y') g_B(x-x',y-y') dx'dy',
\end{equation}
where the function $g_B$ is given by
\begin{equation}\label{gb}
    g_B(x,y)=\int_{-\infty}^{\infty}\frac{\rho_z(z')dz'}{\sqrt{x^2+y^2+z'^2}}.
\end{equation}
Thus, the potential $\Phi$ on the galactic plane is the
convolution of the functions $\Sigma$ and $g_B$. We
can again calculate it applying the convolution theorem
$\Phi=\mathcal{F}^{-1}\left[\mathcal{F}(\Sigma)\;\mathcal{F}(g_B)\right]$,
since we know the surface density $\Sigma(x_i,y_j)$ at every grid
point $(i,j)$.

\begin{figure}
\begin{center}
\includegraphics[width=8.4cm]{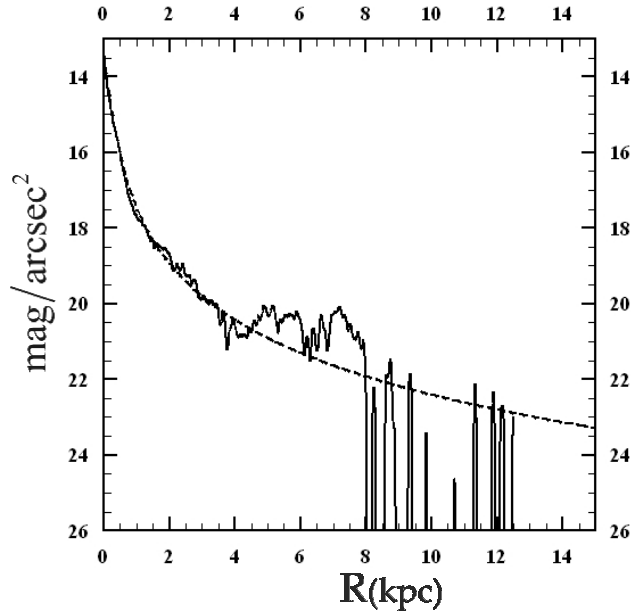}
\end{center}
\caption{The surface brightness of the MHP term in model C. The
solid line represents the mean (over azimuth) surface brightness
after the subtraction of the exponential disc surface brightness
(see Fig.~\ref{fig04}b) from the total one. This part is fitted by
the MHP with the parameter values shown in Table \ref{tabt000}
(dashed curve).} \label{fig08}
\end{figure}

We note that \citet{ls02} and \citet{lsbv04} introduced an
improved version of the Quillen et al (1994) method, where the
image pixels are first smoothed by calculating the azimuthal
Fourier decompositions of the surface density in different radial
zones in polar coordinates. This approach has some advantages,
like a possibility to use more noisy images and to apply radially
non-constant vertical models. However, for high quality images and
for strongly barred galaxies (like our data for NGC 1300) the
differences in the resulting potential should be negligible
regardless of the integration method.

Working in the same way as in model A we obtain expressions
mathematically similar to \eqref{potrthfour} for the potential
corresponding to the luminous component and to \eqref{pottot} for
the total potential. Table 3 in Appendix contains all the values of
the amplitudes for $\Phi_0(R_i), \Phi_{kc}(R_i), \Phi_{ks}(R_i)$
calculated for 101 rings at $R_i$. The specific values for $\Phi_0,
\Phi_{kc}, \Phi_{ks}$ correspond to $6.5\times 10^{10} M_{\odot}$
for the mass of the luminous matter.

Finally, Fig.~\ref{rotcur_3Ddisk} presents the rotation curve
corresponding to model B together with the observational data
\citep{linetal1997} as we did in Fig.~\ref{rotcur_2D} for the 2D
case. The dotted line gives the rotation curve along the major
axis of the bar when we take into account only the luminous
component. We observe that this rotation curve reaches lower
values than the observational one also at the central parts. For
this reason we add a CMC term which is $\approx 3$ times more
massive than the one we added in model A, i.e. now it is
$2.3\times 10^{9} M_{\odot}$. We add also a DH term in order to
get a better description at the outer parts of the disc. The
dashed line gives the rotation curve of the total potential
\eqref{pottot} with the corresponding parameter values given in
Table~\ref{tabt000}.

\subsection{Model C}
In this case we consider that the main part of the bar can be
described with a spherical component. This part could be vaguely
described as ``the bar without its ansae" (Fig.~\ref{fig01}b). For
this purpose we use the modified Hubble profile (MHP)
\citep[see][p. 68]{bt08}
\begin{equation}\label{mhp}
    \rho_{MHP}=j_0\left[1+\left(\frac{r}{a_{MHP}}\right)^2\right]^{-3/2}.
\end{equation}
The parameter $a_{MHP}$ is the scale length and the parameter
$j_0$ determines the mass inside a specific radius (e.g. $r\leq
a_{MHP}$). Note that the mass inside a radius, in MHP model,
diverges logarithmically at large radii and therefore it is
meaningless to refer to total mass.
The corresponding potential via Poisson equation is
\begin{equation}\label{potmhp}
    \Phi_{MHP}=-\frac{G M_h(r)}{r}-\frac{4 \pi G j_0 a_{MHP}^2}{\sqrt{1+\left(\frac{r}{a_{MHP}}\right)^2}},
\end{equation}
where
\begin{multline}\label{mh}
    M_h(r)=4 \pi a_{MHP}^3 j_0 \times\\
    \left[\ln\left(\frac{r}{a_{MHP}}+\sqrt{1+\left(\frac{r}{a_{MHP}}\right)^2}\right)
    -\frac{r}{a_{MHP}}\left[1+\left(\frac{r}{a_{MHP}}\right)^2\right]^{-1/2}\right].
\end{multline}

\begin{figure}
\begin{center}
\includegraphics[width=8.4cm]{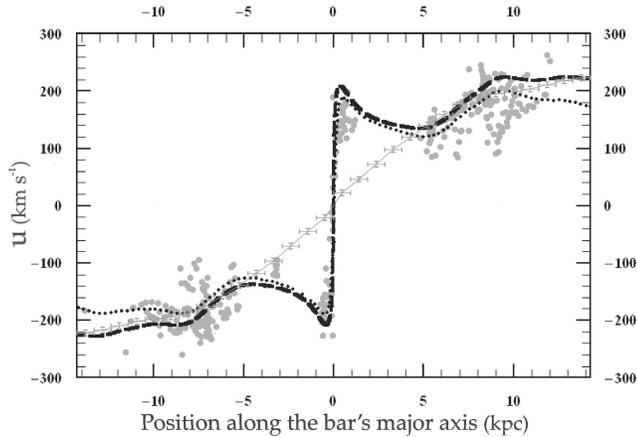}
\end{center}
\caption{Rotation curve together with observed velocities for
model C. Drawn curves are as in Figs.~\ref{rotcur_2D} and
\ref{rotcur_3Ddisk}. The rotation curve of the luminous matter along the bars'
major axis is
given with the dotted curve. The corresponding curve of the total
potential including CMC and DH components,
with the parameter values given in Table \ref{tabt000} is represented by the
dashed curve.} \label{rotcur_3Dbulgedisk}
\end{figure}

By integrating \eqref{mhp} over $z$ we take the expression for the
surface density for MHP, which is
\begin{equation}\label{sdmhp}
    \Sigma_{MHP}(R)=\frac{2j_0 a_{MHP}^3}{a_{MHP}^2+R^2}
\end{equation}
In order to find the values of the best fitting parameters of the
MHP model we subtract the exponential disc $\Sigma_{d}\propto
e^{-R/h_R}$ with $h_R$=10~kpc (see Fig.~\ref{fig04}b) from the
observational surface density $\Sigma_{obs}$ so that the remaining
luminosity at the central region ($R<4$~kpc) is mainly due to the
assumed spherical component. In Fig.~\ref{fig08} we plot the
azimuthally mean surface brightness of the central area of the
galaxy (solid line) after the subtraction of the exponential disc.
We also give the curve corresponding to \eqref{sdmhp} (dashed
curve) with the parameter values shown in Table \ref{tabt000}. We
observe that the MHP gives a reasonable fitting of the inner
3.5-4~kpc.

The total potential corresponding to the luminous mass is the sum of
the MHP component \eqref{potmhp} and the component
corresponding to the remaining mass. The surface density of the
latter component $\Sigma_D$ is derived by subtracting the surface
density $\Sigma_{MHP}$ from the surface density $\Sigma_{obs}$ we
derive from the observations. Thus, the surface density (of the
remaining disc) $\Sigma_D(x_i,y_j)$ at each pixel $(i,j)$ is given
by
\begin{equation}\label{sdc}
    \Sigma_D(x_i,y_j)=\Sigma_{obs}(x_i,y_j)-<\Sigma_{MHP}(x_i,y_j)>,
\end{equation}
where $<\Sigma_{MHP}(x_i,y_j)>$ is the mean surface density
$\Sigma_{MHP}$ inside the area of the squared pixel centered at
$(x_i,y_j)$. We consider this component as a 3D exponential disc
with vertical distribution given by \eqref{kruitmodel}. In order to obtain
an expression similar to \eqref{potrthfour} for the disc in model C
we work in the same way as in model B. Table 4 in Appendix gives the
values of the amplitudes of $\Phi_0(R_i), \Phi_{kc}(R_i),
\Phi_{ks}(R_i)$ of the potential corresponding to the disc
component.

Figure~\ref{rotcur_3Dbulgedisk} shows the data of the
observational rotational velocities \citep{linetal1997} as in the
two previous models together with the rotation curve of the
luminous component (both MHP and disc) along the major axis of the
bar represented by the dotted curve. Also in this case the
activation of the two additional potential terms, especially the
DH one, allows a better relation between the rotation curve of the
model along its bars major axis and the inclination of the
observed \Hi curve at large distances. The values of their
parameters for model C are also shown in Table \ref{tabt000}.

\begin{figure*}
\begin{center}
\includegraphics[width=14.5cm]{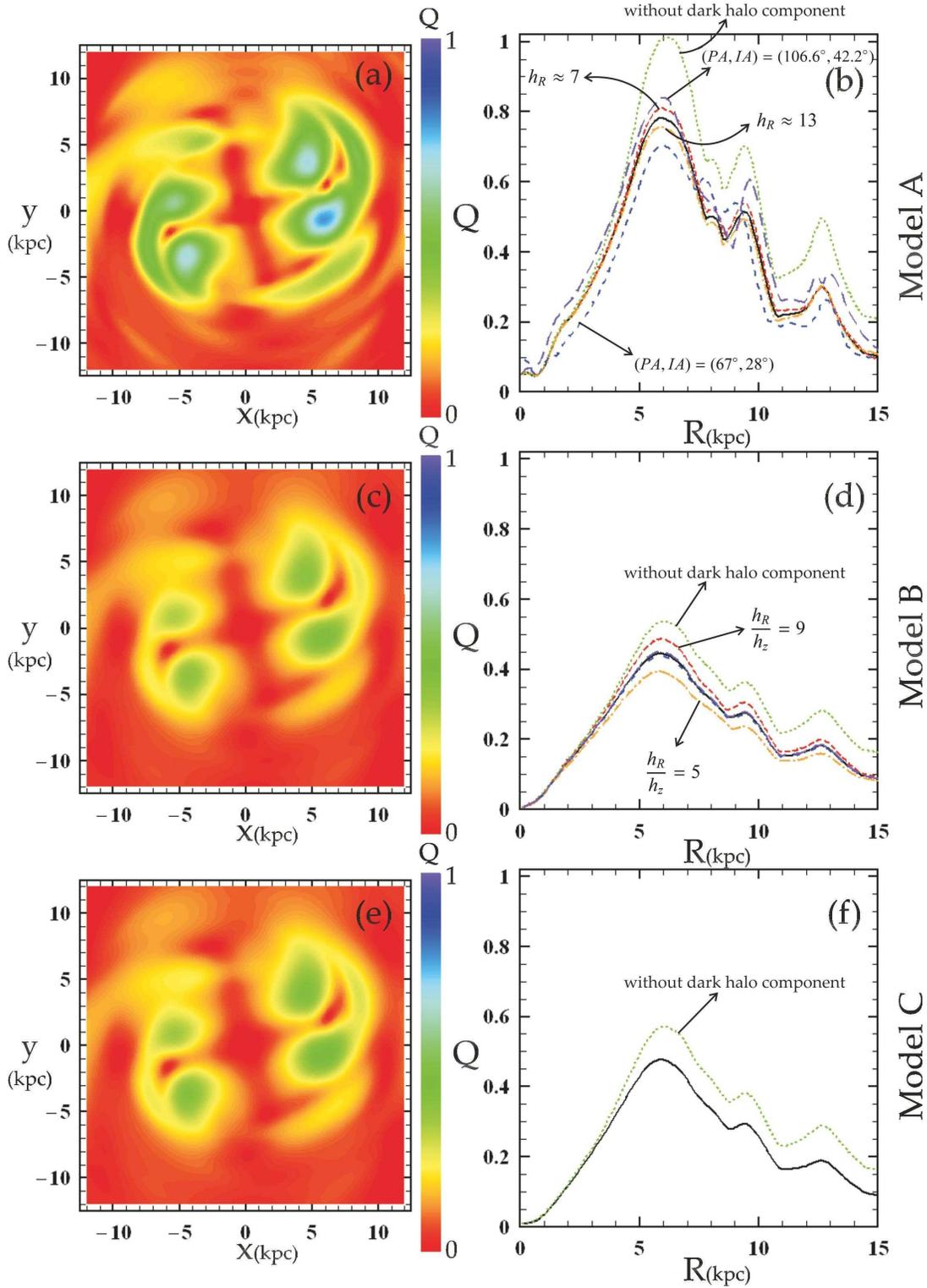}
\end{center}
\caption{The $Q$ variation in our three models for several sets of
the potential parameters. The panels of the left-hand column give
$Q$ for models ``A'' (a), ``B''(c) and ``C''(e). The panels of the
right-hand column give the maximum $Q$ as a function of $R$ (black
solid curves). Additional curves refer to the radial variation of
maximum $Q$ in models with different parameters than those used in
the basic set-up of the three general models. For model ``A'' in
panel (b) the dotted green curve refers to a model without dark
halo, the magenta and blue dashed curves to different orientation
parameters, while the dashed pink and dark yellow ones to models
with disc scale lengths $h_R=7$~kpc and $h_R=13$~kpc respectively.
Arrows help to better trace each curve. For model ``B'' in panel
(d), additional curves are given for a ``model B'' without dark
halo component (dotted green curve) and models with $h_R/h_z$
=5,9. Finally for model ``C'', panel (f), an additional curve
gives the radial variation of $Q$ again in a model that with
respect to ``C'' does not have any dark halo component.}
\label{qforces}
\end{figure*}


\begin{figure*}
\begin{center}
\includegraphics[width=15cm]{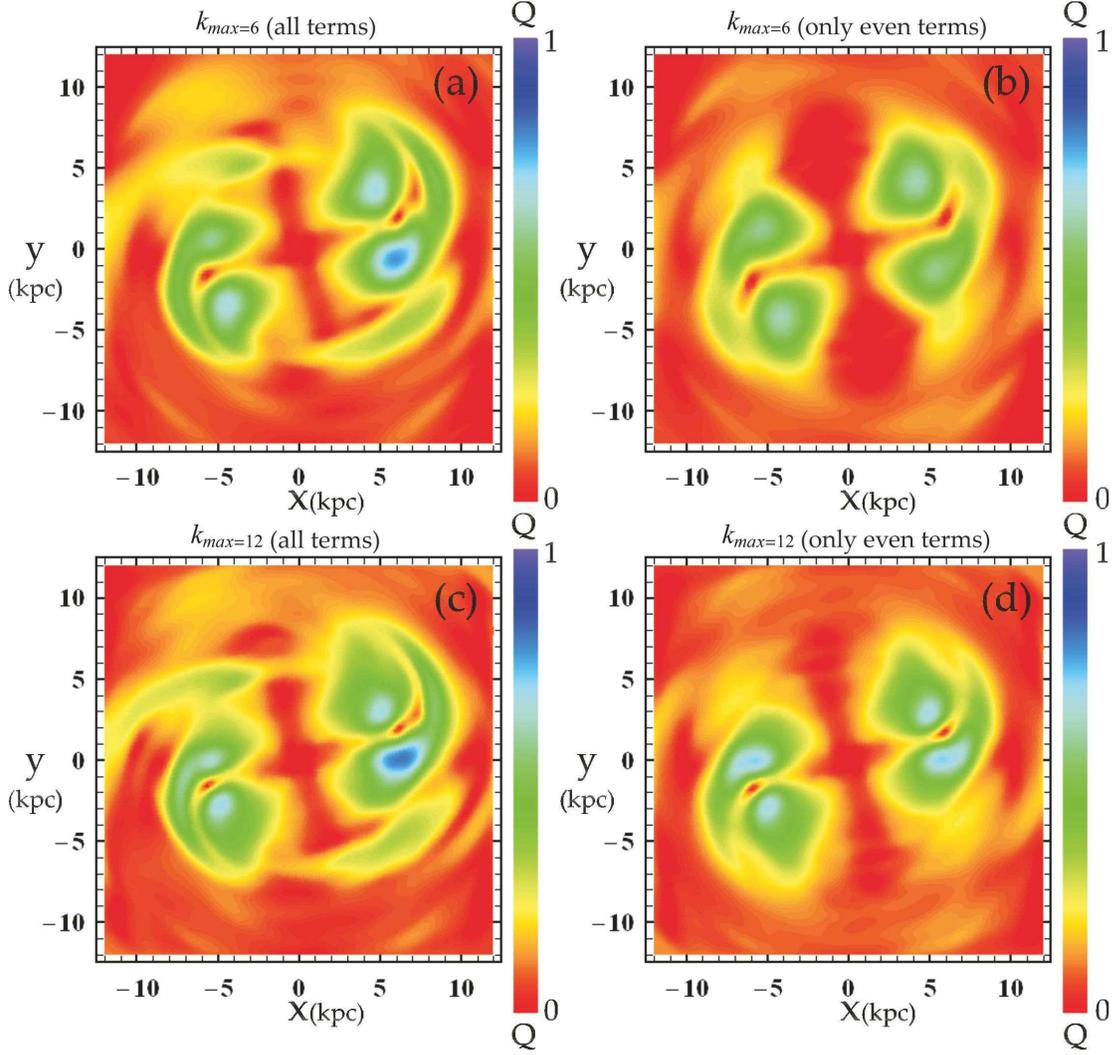}
\end{center}
\caption{The $Q$ value maps for model A on the galactic plane for
$k_{max}=6$, $k_{max}=12$ (first and second row respectively) with
and without odd terms (left-hand column and right hand column
respectively). A combination that can be used for reliable
dynamical models is with $k_{max}=6$, including all terms (panel
(a)).} \label{figqm}
\end{figure*}

\begin{figure*}
\begin{center}
\includegraphics[width=15.5cm]{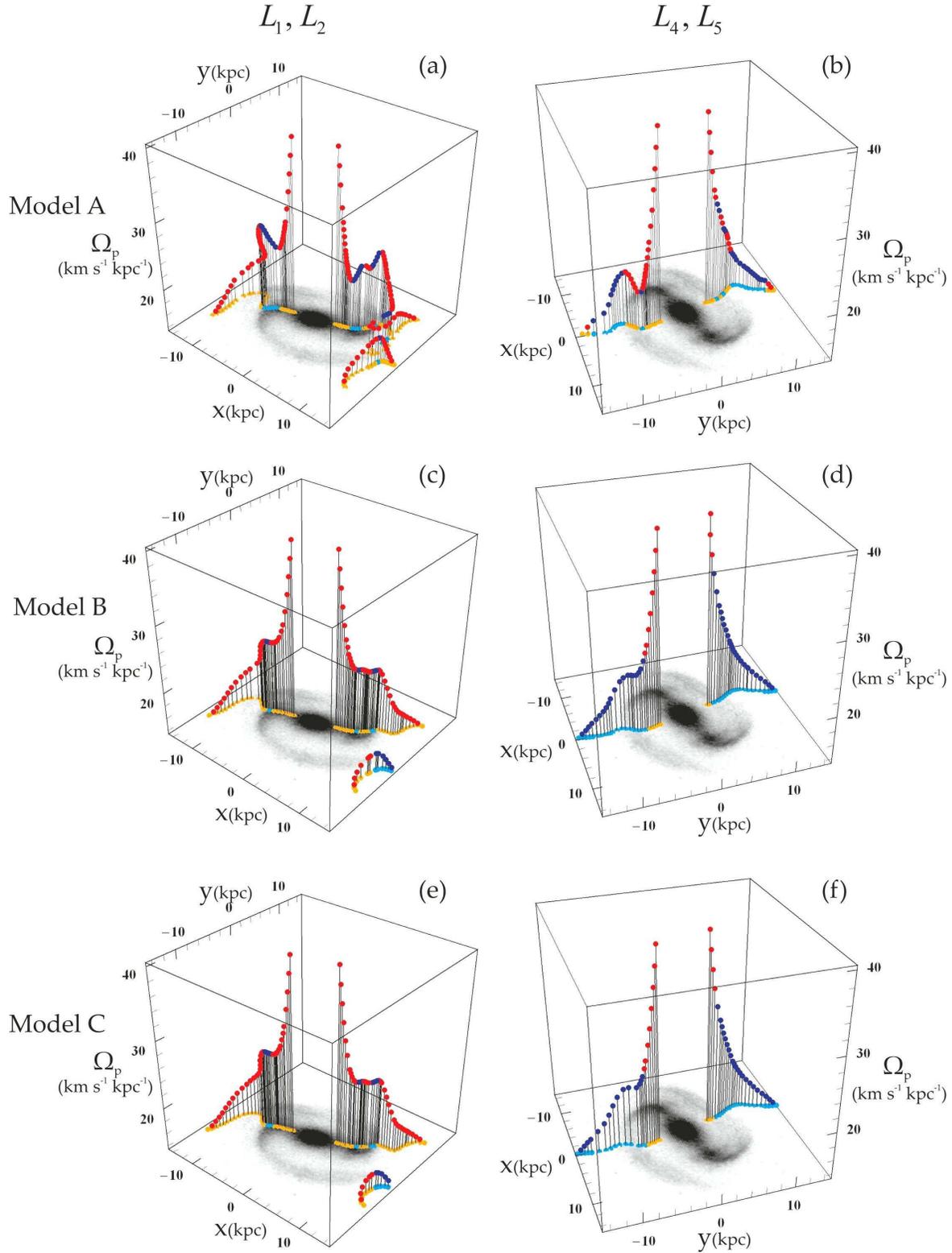}
\end{center}
\caption{The loci of the Lagrangian points ($L_1, L_2$ in the
left-hand column and $L_4, L_5$ in the right-hand column) for all
the models as a function of the $\Omega_p$. The red (blue) points
correspond to unstable (stable) Lagrangian points. In each panel we
have projected the Lagrangian points on the galactic plane. We
observe that in all models there are ranges of $\Omega_p$ values for
which we have multiple Lagrangian points. This property is more
evident in model A, where we observe longer ranges of $\Omega_p$
values corresponding to multiple Lagrangian points. Note that the
separated group of points which are located at high $x$ values (seen
in all the left-hand column panels) correspond to multiple $L_1$
Lagrangian points ordered near-parallel to the $y$ axes and they
refer to cases with low $\Omega_p$ values.} \label{lagrpoints}
\end{figure*}

\begin{figure*}
\begin{center}
\includegraphics[width=\textwidth]{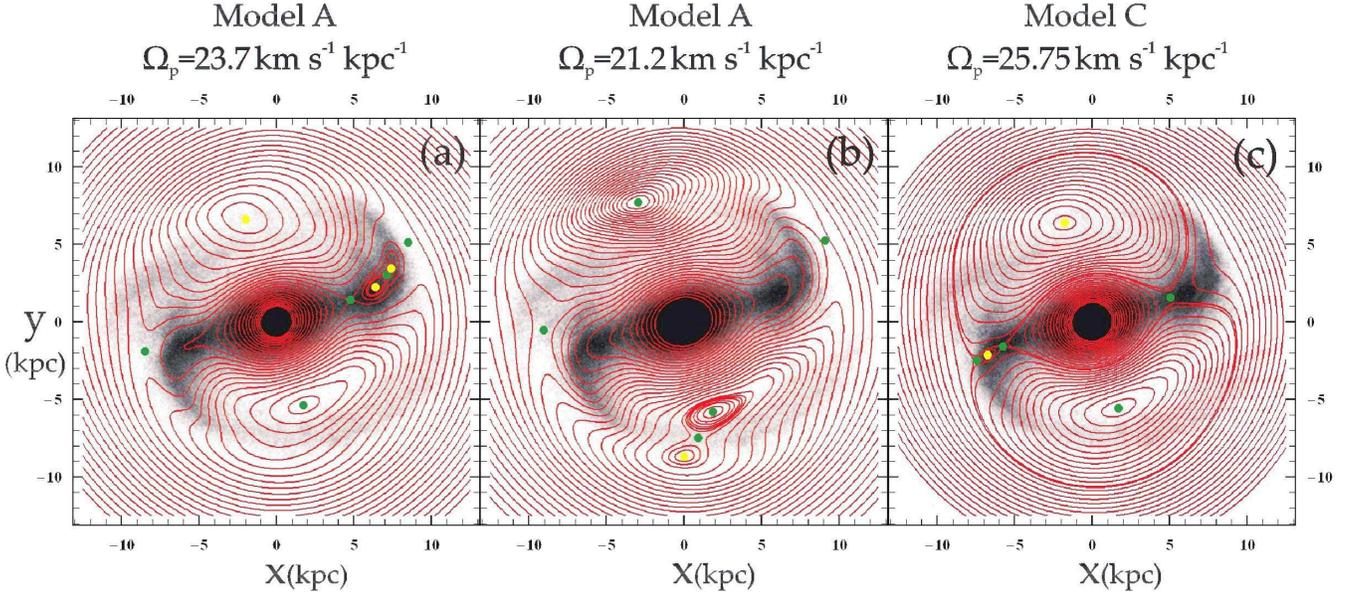}
\end{center}
\caption{Some characteristic examples of multiple Lagrangian
points. We plot the contour lines of the effective potential
overlapped with the image of the galaxy for the cases and the
$\Omega_p$ values indicated in the figure. Big yellow (green) dots
correspond to stable (unstable) Lagrangian points. In (a) we see 3
unstable and 2 stable $L_1$ points. In (b) we have 2 stable and 1
unstable $L_5$ points. In (c) finally we see 2 unstable and 1
stable $L_2$ points. Note that the considered sense of rotation is
clockwise.} \label{contlns}
\end{figure*}

\section{The Nonlinearity of the models}
\label{sec:forces} As a measure of the nonlinearity of the models
we consider the quantity
\begin{equation}
\label{qquant}
\begin{gathered}
    Q=\frac{|\mathbf{\Delta
    F}|}{|\mathbf{F_{axis}}|}=\frac{|\mathbf{F_{total}-F_{axis}}|}{|\mathbf{F_{axis}}|}=
\\
    \dfrac{\sqrt{\left(\dfrac{\partial \Phi_{total}}{\partial R}-\dfrac{\partial \Phi_{axis}}{\partial R}\right)^2+
    \left(\dfrac{1}{R}\dfrac{\partial \Phi_{total}}{\partial \theta}\right)^2}}
    {\left|\dfrac{\partial \Phi_{axis}}{\partial R}\right|}
\end{gathered}
\end{equation}
which is the ratio of the modulus of the vectorial difference of
the total force (axisymmetric and non-axisymmetric) from the
axisymmetric force, over the modulus of the axisymmetric force.
Thus, $Q$ measures the non-axisymmetric force perturbation,
normalized over the axisymmetric one. This parameter
describes well the non-linear effect of the non-axisymmetric
components but it is unable to determine by itself the global
dynamics in a rotating system (see below the discussion about the
importance of $\Omega_p$).

In the past, the strength of bars and spirals in disc galaxies has
been quantified by means of similar indices
\citep[e.g.][]{bb01,v06}. The index introduced by \citet{bb01} has
been widely used in large samples of galaxies \citep{ls02,
bbkeep04, lsbv04}. This index is similar to the quantity $Q$
defined in Eq.~\eqref{qquant}. The variation of $Q$ in our models
is presented in Fig.~\ref{qforces}.

At the left-hand column the panels show in color diagrams the
strength of the non-axisymmetric force, measured by $Q$, for the
three models. Each model is indicated at the right edge of the
figure. At the right-hand column we plot the maximum $Q$ as a
function of $R$ (black solid curves). We observe that the overall
maximum perturbation strength reaches higher values ($\approx
0.8$) in model A (pure 2D), while in the other two models it is
lower reaching $\approx 0.45$ and $\approx 0.5$ for models B and C
respectively. The maximum $Q$ values of models B and C are close
to those presented by \citet{ls02}, while the pure 2D case has a
significant larger overall maximum $Q$ value. In each panel of the
right-hand column we have plotted additional maximum $Q$ curves
corresponding to different parameter values, which are indicated
in the figure. The green curves give the maximum $Q$ profiles
without the DH component. We observe that the maximum $Q$ values,
with the DH component included, are approximately 0.8-0.85 of the
corresponding values neglecting the DH component. Such values are
within the range of values given by \citet{bls04}, although
towards the lower limit. We studied also the robustness of the
maximum $Q$ profile to the adopted orientation parameters. We
considered two additional orientation parameter sets i.e. (PA, IA)
= $(106.6^\circ, 42.2^\circ)$ (dashed magenta curve,
Fig.~\ref{qforces}b) and (PA, IA) = $(67^\circ, 28^\circ)$ (dashed
blue curve, Fig.~\ref{qforces}b). The first set of the orientation
parameters is the one obtained from near-infrared data by
\citet{gpp2004}. The second set has angles deviating from the
values we finally used as much as the first one, but to the
opposite direction. We observe, that the maximum $Q$ value, in
these cases, is affected by less than 10\%. We also checked the
dependence of the maximum $Q$ value on the changes of the
zero-value level, which yield to disc scale lengths of 7~kpc
(dashed red curve, Fig.~\ref{qforces}b) and 13~kpc (dashed-dotted
orange curve, Fig.~\ref{qforces}b), respectively. Note, that the
exponential discs with these scale lengths do not fit the
corresponding data as well as in the original $h_R$=10~kpc case.
We observe, that the maximum $Q$ value, in these cases, is
affected by less than 5\%.

In Fig.~\ref{qforces}d we see that the maximum $Q$ value, in model
B, varies approximately by 10\% when the ratio $h_R/h_z$ varies
$1\sigma$ from its mean value (see dashed-dotted orange and dotted
red curves in Fig.~\ref{qforces}d). We also certified, that the
$Q$ profiles almost coincide within $1\sigma$ variation of the
exponent $n$ of Eq.~\eqref{kruitmodel} from its mean value.

Special attention has been given to the number and kind (even or
odd) of the Fourier terms to be included in the potential. The
morphological asymmetries observed in the galaxy underline the
importance of the inclusion of the odd terms for dynamical
studies. The number and the kind of the Fourier terms may affect
the maximum $Q$ value as well as the 2D distribution of the $Q$
value on the galactic plane. In order to study these effects we
constructed $Q$ value maps on the galactic plane for different
combinations of the Fourier terms (Fig.~\ref{figqm}). We present
the results for model A since it is the model with the highest
non-axisymmetric amplitudes in the Fourier decomposition.
The upper two panels (a,b) of Fig.~\ref{figqm} include terms up to
$k_{max}=6$ in Eq.~\eqref{potrthfour}, while the two panels at the
bottom (c,d) up to $k_{max}=12$. The left panels (a,c) include
\textit{all} $k$ terms, while the right ones (b,d) only the even
terms, as indicated at the top of each of them. By including all
terms, the maximum $Q$ value for the case we consider up to
$k_{max}=6$ is 0.96 of the maximum value corresponding to
$k_{max}=12$. By including higher order terms $(k_{max}>12)$, the
maximum $Q$ value does not get further refinement. On the other
hand, by neglecting the odd terms, the $Q$ map clearly fails to
reproduce all the asymmetries and the maximum $Q$ value becomes at
least 10\% lower than that with the odd terms even for the
$k_{max}=12$ case. More specifically, the maximum $Q$
values corresponding to panels (a), (b), (c) of Figure \ref{figqm}
are 0.96, 0.89, 0.84 of the maximum $Q$ value we get in
Fig.~\ref{figqm}c. We can observe, that at the sides of the bar
the prominent green "tails",  which essentially describe the force
field at the spirals' region, are conspicuous only in panels (a)
and (c). Also at radii around 8 and 9 kpc, at the ends of the bar,
only panels (a) and (c) give a rather constant green shade, while
at the same radii green color fades out in panel (b), but also in
(d). Evident is that in terms of nonlinearity of the potential
the odd terms are essential, while including $k>6$ terms does not
improve the models at a level worth the additional computing time
needed for the calculations. Thus, a reliable combination that can
be used in dynamical studies is $k_{max}=6$ with all even and odd
terms included (Fig.~\ref{figqm}a).

In all examined cases the forcing
of the models suggests that our systems are strongly nonlinear,
since the perturbing term is at least higher than 40\% of the
axisymmetric background. This is a usual situation in models of
strong barred galaxies \citep{kc1996}.

We note that potential evaluations for NGC 1300 have been
previously done by \citet{e89}, \citet{h90}, \citet{lk96},
\citet{apvm01} and more recently by \citet{ls02} using a 2MASS
K-band image and by \citet{lsbv04} by means of OSUBGSG H-band
data. These potentials have some features similar with our general
model (for a discussion see Sect.~7). In our case, emphasis is
given in constructing a potential function suitable for dynamical
studies. We kept fixed the quantities that could be derived
directly from our near-infrared K-band data, while the parameters
that could be only confined within some range have been treated as
free parameters. However, the most important parameter for the
dynamics of a rotating disc galaxy is the pattern speed \citep[for
a review see][]{gc02}, which is an unknown quantity. The most
simple assumption is that the system we are studying has a unique
pattern speed. We take it as a basic hypothesis, the validity of
which will be evaluated from the feedback we will get from
response models (to be presented in subsequent papers). At any
rate, it is the \textit{effective potential} that determines the
force field in our calculations. The dynamics of two systems with
the same potential $\Phi$ but with considerably different pattern
speeds are considerably different \citep[see e.g.][]{pcg91}. An
investigation of the basic categories of effective potentials is
presented in Sect.~6 below.


\section{Effective potentials}
The locations of the stationary (stable and unstable) Lagrangian
points strongly affect the dynamical properties of a rotating
galaxy. The Lagrangian points and their stability structure the
phase space at the corotation region. Given a mass distribution, the
positions of the Lagrangian points are determined by the value of
the pattern speed $\Omega_{p}$.
In the rotating frame we have the
conservation of the Jacobi integral
\begin{equation}\label{ejac}
    E_{j}=\frac{1}{2}v^2+\Phi_{eff},
\end{equation}
where $\Phi_{eff}$ is the effective potential
\begin{equation}\label{phieff}
    \Phi_{eff}(x,y)=\Phi_{T}(x,y)-\frac{1}{2}\Omega_p^2(x^2+y^2).
\end{equation}
The stationary Lagrangian points at the equilibrium positions are defined by
\begin{equation}\label{lagrcond}
    \frac{\partial \Phi_{eff}(x,y)}{\partial x}=\frac{\partial
      \Phi_{eff}(x,y)}{\partial
    y}=0.
\end{equation}

This is the case of course, where the system rotates with a single
pattern speed. In systems with multiple pattern speeds, multiple
corotation radii will be defined on the disc. If bar and spirals
rotate with different $\Omega_p$, each component will have its own
corotation radius. In what follows we take the single pattern
speed case as our initial assumption. Its validity and its
limitation will be examined by means of models in subsequent
papers.

Fig.~\ref{lagrpoints} shows in a compact way how the number, the
location and the stability of the Lagrangian points change on the
disc for the various effective potentials. In
Fig.~\ref{lagrpoints}a we give the positions $(x,y)$ of the $L_1,
L_2$ Lagrangian points as a function of $\Omega_{p}$ for model A.
The calculation of the stability of the equilibrium points is done
by following the standard procedure that can be found in relevant
textbooks   \citep[see e.g.][pp. 179-183]{bt08}. The red and blue
points correspond to the unstable and stable Lagrangian points,
respectively. At the basis of the cubes in all panels it is
plotted the surface density of the galaxy together with the
projections of the corresponding Lagrangian points. The projected
orange and light blue points correspond to the unstable and stable
Lagrangian points, respectively. Figure \ref{lagrpoints}b is
similar to Fig.\ref{lagrpoints}a and shows the positions of $L_4,
L_5$ Lagrangian points in model A. We observe that there are
ranges of $\Omega_{p}$ values for which there are more than one
$L_1, L_2, L_5$ points. Figures \ref{lagrpoints}c,d and
\ref{lagrpoints}e,f are similar to Fig.~\ref{lagrpoints}a,b but
for the model B and C, respectively. We see that the distribution
of Lagrangian points have a very similar behavior in models B and
C. Moreover, the variability of the positions of the Lagrangian
points with $\Omega_{p}$ seems smoother in models B and C.
However, even in these cases we have ranges of $\Omega_{p}$ values
corresponding to multiple Lagrangian points (see
Fig.~\ref{contlns}c).

Figure \ref{contlns} presents some characteristic cases of
effective potentials with multiple Lagrangian points. Stable
(unstable) Lagrangian points are plotted with yellow (green) big
dots. Fig.~\ref{contlns}a shows the isocontours of the effective
(total) potential, in model A, for $\Omega_{p}=23.7\text{km
sec}^{-1}\text{kpc}^{-1}$. In this figure we observe 5 $L_1$
Lagrangian points (3 unstable and 2 stable) at the region close to
the end of the bar. Figure \ref{contlns}b corresponds to the same
model A but for $\Omega_{p}=21.2\text{km
sec}^{-1}\text{kpc}^{-1}$. We observe 3 $L_5$ Lagrangian points (2
stable, 1 unstable). Finally, Fig.~\ref{contlns}c corresponds to
model C for $\Omega_{p}=25.75\text{km sec}^{-1}\text{kpc}^{-1}$
and shows 3 $L_2$ Lagrangian points (2 unstable, 1 stable).

\section{Discussion and Conclusions}

In this paper, we propose three different general models for the
potential of the grand design barred-spiral galaxy NGC~1300, based
on observations in the near-infrared. The ultimate goal of our
work is the investigation of the stellar dynamics of this galaxy
by means of response models and orbital theory. More precisely we
want to reveal the dynamical mechanisms that shape the observed
structure of the galaxy. This is done in the forthcoming papers of
this series.

Observations in the near-infrared provide the most reliable
potential models of disc galaxies, since they are excellent
indicators of the distribution of mass associated to stellar
matter. The needed additional assumptions refer mainly to the
distribution of luminous matter in the third dimension (outside
the galactic plane) and to the distribution of the non-luminous
matter (dark halo and central mass concentration).

The three proposed models are: The pure 2D case (Model A), where
all luminous matter is considered on the galactic plane; the thick
disc case (Model B) where all luminous matter is considered in a
3D disc with a constant scale height; and the ``spheroidal case''
(Model C) in which the major part of the bar is a spherical
modified Hubble profile. The remaining luminous matter is
attributed to a 3D disc as in Model B. In all these cases the
calculation of the potential is performed by means of an FFT
transform technique, that has been introduced in various versions
by \citep[][]{hohhoc1969,quietal1994,ls02}.

These three general models can be considered as limiting cases in
the parameter space of models whose responses and orbital content
will be studied. Let us consider an unknown parameter, e.g. the
disc thickness. If we find that we obtain better feedback in Model
A than in Model B, we will conclude that we have more accurate
dynamics closer to the thin disc limit. Only the study of the
dynamics in a large number of effective potentials will show which
configuration compares best with the observed morphology.

As we have seen in Section 5, all models are strongly nonlinear.
The relative force perturbation (regarding the axisymmetric
background) in all models is high reaching to 45\%-50\% in Models
B and C and to even higher value 80\% in Model A. We note that
\textit{for normal (non-barred) spiral} galaxies the models that
reproduce successfully the morphology of this type of galaxies
have maximum force perturbations typically of the order of 5-10\%
\citep{pcg91, ls02, lsbv04, v06} and they are already
characterized by nonlinear effects. Nonlinear effects are even
stronger in larger perturbation characteristic of bars
\citep[see][and others]{kc1996, ls02, lsbv04, bbkeep04}. Thus, it
is expected that all models will present a strongly nonlinear
behavior with extended chaotic regions. Note that we certified the
robustness of the force perturbation in each model to small
changes of the various required parameter values (e.g. thickness
of the disc).

The nonlinearity of a model, expressed as the relative force
perturbation, is an important parameter for studying its dynamics,
since it refers to the total distribution of matter (luminous and
dark) in the galaxy. However, the estimation of the potential of a
disc galaxy alone does not determine its dynamics. The main, and
most important, parameter is the pattern speed and this cannot be
unambiguously determined from observations. The Coriolis forces
totally alter the landscape of the gravitational field, as we can
see in the various effective potentials. A potential by itself can
tell us whether or not nonlinear phenomena can play an important
role in the dynamics of the system. However, for the same
potential we can have totally different dynamics. By varying the
pattern speed the equilibrium (Lagrangian) points are shifted and
may alter their stability character. In our analysis we find cases
with multiple Lagrangian points. This indicates that in real
galaxies the existence of multiple Lagrangian points can be a
common phenomenon. Our models give the opportunity to study the
dynamics of such realistic systems, something that has not been
done yet. The effect of the variation of the pattern speed on the
dynamics of NGC~1300 will be presented in the forthcoming papers
of this series.

\section*{Acknowledgments}
We thank Prof. G.~Contopoulos for fruitful discussions. P.A.P thanks
ESO for a two-months stay in Garching as visitor, where part of this
work has been completed. We would also like to thank the anonymous
referee for constructive comments that helped the clarity and
the presentation of our paper.

\bibliographystyle{mn2e}
\bibliography{evk}

\begin{thebibliography}{}

\bibitem[\protect\citeauthoryear{{Aguerri}, {Prieto}, {Varela} \&
  {Munoz-Tunon}}{{Aguerri} et~al.}{2001}]{apvm01}
{Aguerri} J.,  {Prieto} M.,  {Varela} A.,    {Munoz-Tunon} C.,  2001, Astroph.
  S. Sc., 276, 611

\bibitem[\protect\citeauthoryear{{Binney} \& {Tremaine}}{{Binney} \&
  {Tremaine}}{2008}]{bt08}
{Binney} J.,  {Tremaine} S.,  2008, Galactic Dynamics.
Princeton University Press, New Jersey

\bibitem[\protect\citeauthoryear{{Block}, {Buta} \& {Knapen}}{{Block}
  et~al.}{2004}]{bbkeep04}
{Block} D.,  {Buta} R.,    {Knapen} J. e.~a.,  2004, \aj, 128, 183

\bibitem[\protect\citeauthoryear{{Buta} \& {Block}}{{Buta} \&
  {Block}}{2001}]{bb01}
{Buta} R.,  {Block} D.~L.,  2001, \apj, 550, 243

\bibitem[\protect\citeauthoryear{{Buta}, {Laurikainen} \& {Salo}}{{Buta}
  et~al.}{2004}]{bls04}
{Buta} R.,  {Laurikainen} E.,    {Salo} H.,  2004, \aj, 127, 279

\bibitem[\protect\citeauthoryear{{Contopoulos}}{{Contopoulos}}{2004}]{gc02}
{Contopoulos} G.,  2004, Order and Chaos in Dynamical Astronomy.
Springer-Verlag, {New York}

\bibitem[\protect\citeauthoryear{Contopoulos \& Patsis}{Contopoulos \&
  Patsis}{2008}]{cia2008}
Contopoulos G.,  Patsis P.~A.,  2008, in Contopoulos G.,  Patsis P.~A.,  eds,
  Chaos in Astronomy Lecture Notes In Physics.
Springer-Verlag, Berlin

\bibitem[\protect\citeauthoryear{de Grijs}{de~Grijs}{1998}]{degrijs1998}
de Grijs R.,  1998, \mnras, 299, 595

\bibitem[\protect\citeauthoryear{de Vaucouleurs, de Vaucouleurs, Corwin, Buta,
  Paturel \& Fouque}{de~Vaucouleurs et~al.}{1991}]{devacetal1991}
de Vaucouleurs G.,  de Vaucouleurs A.,  Corwin H.~G.,  Buta R.,  Paturel G.,
  Fouque P.,  1991, Third Reference Catalogue of Bright Galaxies, 363

\bibitem[\protect\citeauthoryear{{Elmegreen}, {Elmegreen}, {Chromey} \&
  {Hasselbacher}}{{Elmegreen} et~al.}{1996}]{eech96}
{Elmegreen} D.~M.,  {Elmegreen} B.~G.,  {Chromey} F.~R.,    {Hasselbacher}
  D.~A.,  1996, \apj, 469, 131

\bibitem[\protect\citeauthoryear{{England}}{{England}}{1989}]{e89}
{England} M.~N.,  1989, \apj, 337, 191

\bibitem[\protect\citeauthoryear{{Fabbiano}, {Gioia} \&
  {Trinchieri}}{{Fabbiano} et~al.}{1988}]{fgt88}
{Fabbiano} G.,  {Gioia} I.~M.,    {Trinchieri} G.,  1988, \apj, 324, 749

\bibitem[\protect\citeauthoryear{Gaddoti}{Gaddoti}{2008}]{gad2008}
Gaddoti D.~A.,  2008, \mnras, 384, 420

\bibitem[\protect\citeauthoryear{Grosb{\o}l}{Grosb{\o}l}{2008}]{gro2008}
Grosb{\o}l P.,  2008, in Contopoulos G.,  Patsis P.~A.,  eds, Chaos in
  Astronomy Lecture Notes In Physics.
Springer-Verlag, Berlin, p.~23

\bibitem[\protect\citeauthoryear{{Grosb{\o}l} \& {Dottori}}{{Grosb{\o}l} \&
  {Dottori}}{2008}]{gd08}
{Grosb{\o}l} P.,  {Dottori} H.,  2008, \aa, 490, 87

\bibitem[\protect\citeauthoryear{Grosb{\o}l \& Patsis}{Grosb{\o}l \&
  Patsis}{1998}]{gropat1998}
Grosb{\o}l P.,  Patsis P.~A.,  1998, \aa, 336, 840

\bibitem[\protect\citeauthoryear{Grosb{\o}l, Patsis \& Pompei}{Grosb{\o}l
  et~al.}{2004}]{gpp2004}
Grosb{\o}l P.,  Patsis P.~A.,    Pompei E.,  2004, \aa, 423, 849

\bibitem[\protect\citeauthoryear{Hohl \& Hockney}{Hohl \&
  Hockney}{1969}]{hohhoc1969}
Hohl F.,  Hockney R.~W.,  1969, J. Comput. Phys., 4, 306

\bibitem[\protect\citeauthoryear{{Hunter}}{{Hunter}}{1990}]{h90}
{Hunter} J.,  1990, Ann.N.Y. Acad. Sci., 596, 174

\bibitem[\protect\citeauthoryear{{Kaufmann} \& {Contopoulos}}{{Kaufmann} \&
  {Contopoulos}}{1996}]{kc1996}
{Kaufmann} D.~E.,  {Contopoulos} G.,  1996, \aa, 309, 381

\bibitem[\protect\citeauthoryear{Kregel, van~der Kruit,  \& de Grijs}{Kregel
  et~al.}{2002}]{kreetal2002}
Kregel M.,  van~der Kruit P.~C.,     de Grijs R.,  2002, \mnras, 334, 646

\bibitem[\protect\citeauthoryear{{Laurikainen} \& {Salo}}{{Laurikainen} \&
  {Salo}}{2002}]{ls02}
{Laurikainen} E.,  {Salo} H.,  2002, \mnras, 337, 1118

\bibitem[\protect\citeauthoryear{{Laurikainen}, {Salo}, {Buta} \&
  {Vasylyev}}{{Laurikainen} et~al.}{2004}]{lsbv04}
{Laurikainen} E.,  {Salo} H.,  {Buta} R.,    {Vasylyev} S.,  2004, \mnras, 355,
  1251

\bibitem[\protect\citeauthoryear{{Lindblad} \& {Kristen}}{{Lindblad} \&
  {Kristen}}{1996}]{lk96}
{Lindblad} P.,  {Kristen} H.,  1996, \aa, 313, 733

\bibitem[\protect\citeauthoryear{Lindblad, Kristen, Joersaeter \&
  Hoegbom}{Lindblad et~al.}{1997}]{linetal1997}
Lindblad P.~A.~B.,  Kristen H.,  Joersaeter S.,    Hoegbom J.,  1997, \aa, 317,
  36

\bibitem[\protect\citeauthoryear{{Patsis}, {Contopoulos} \& {Grosbol}}{{Patsis}
  et~al.}{1991}]{pcg91}
{Patsis} P.~A.,  {Contopoulos} G.,    {Grosbol} P.,  1991, \aa, 243, 373

\bibitem[\protect\citeauthoryear{{Patsis}, {H{\'e}raudeau} \&
  {Grosb{\o}l}}{{Patsis} et~al.}{2001}]{phg01}
{Patsis} P.~A.,  {H{\'e}raudeau} P.,    {Grosb{\o}l} P.,  2001, \aa, 370, 8

\bibitem[\protect\citeauthoryear{{Prieto}, {Aguerri}, {Varela} \&
  {Mu{\~n}oz-Tu{\~n}{\'o}n}}{{Prieto} et~al.}{2001}]{pea2001}
{Prieto} M.,  {Aguerri} J.~A.~L.,  {Varela} A.~M.,    {Mu{\~n}oz-Tu{\~n}{\'o}n}
  C.,  2001, \aa, 367, 405

\bibitem[\protect\citeauthoryear{Quillen, Frogel \& Gonzalez}{Quillen
  et~al.}{1994}]{quietal1994}
Quillen A.~C.,  Frogel J.~A.,    Gonzalez R.~A.,  1994, \apj, 437, 162

\bibitem[\protect\citeauthoryear{Rix \& Rieke}{Rix \& Rieke}{1993}]{rixrie1993}
Rix H.~W.,  Rieke M.~J.,  1993, \apj, 418, 123

\bibitem[\protect\citeauthoryear{{Seigar}, {Block}, {Puerari}, {Chorney} \&
  {James}}{{Seigar} et~al.}{2005}]{sea2005}
{Seigar} M.~S.,  {Block} D.~L.,  {Puerari} I.,  {Chorney} N.~E.,    {James}
  P.~A.,  2005, \mnras, 359, 1065

\bibitem[\protect\citeauthoryear{van~der Kruit}{van~der
  Kruit}{1988}]{kruit1988}
van~der Kruit P.~C.,  1988, \aa, 192, 117

\bibitem[\protect\citeauthoryear{van~der Kruit}{van~der
  Kruit}{2002}]{kruit2002}
van~der Kruit P.~C.,  2002, in Costa G.~S.~D.,  Jerjen H.,  eds, The Dynamics,
  Structure \& History of Galaxies Vol.~273 of ASP Conference Proceedings.
Astronomical Society of the Pacific, San Francisco, p.~7

\bibitem[\protect\citeauthoryear{{Vorobyov}}{{Vorobyov}}{2006}]{v06}
{Vorobyov} E.~I.,  2006, \mnras, 370, 1046

\end{thebibliography}

\label{lastpage}

\end{document}